\documentclass[
    ,draft            
  ]
  {aipproc}
\layoutstyle{8x11single}

\usepackage{color}
\usepackage{xspace}
\usepackage{geometry}

\newcommand{\superk}    {Super-Kamiokande\xspace}       
\newcommand{\sk}{Super-K\xspace}
\newcommand{\cz}{\ensuremath{\cos \theta_z}\xspace}
\newcommand{\degrees}{\ensuremath{^\circ}}
\newcommand{\dm}[1]{\ensuremath{ \Delta m^{2}_{#1} }\xspace }
\newcommand{\sn}[1]{\ensuremath{ \sin^{2}\! \theta_{#1} }\xspace }
\newcommand{\snt}[1]{\ensuremath{ \sin^{2}\! 2\theta_{#1} }\xspace }
\newcommand{\dmnew}{\dm{\rm{new}}}

\newcommand{\uesq}{\ensuremath{|U_{e4}|^2}\xspace}
\newcommand{\umsq}{\ensuremath{|U_{\mu4}|^2}\xspace}
\newcommand{\utsq}{\ensuremath{|U_{\tau4}|^2}\xspace}

\newcommand{\nue}       {\ensuremath{\nu_{e}}\xspace}
\newcommand{\numu}      {\ensuremath{\nu_{\mu}}\xspace}
\newcommand{\nutau}     {\ensuremath{\nu_{\tau}}\xspace}

\newcommand{\nuebar}{\ensuremath{\overline{\nu}_{e}}\xspace}

\newcommand{\val}[2]{\ensuremath{#1 \; \mathrm{#2}\xspace}}

\newcommand{\fref}[1]{Fig.~\ref{fig:#1}}

\begin{document}

\title{Recent Atmospheric Neutrino Results from \superk}


\author{Alexander Himmel\\ for the Super-Kamiokande Collaboration}{
  address={Department of Physics, Duke University, Durham, North Carolina, USA}
}

\begin{abstract}
The \superk experiment has collected more than 11 live-years of atmospheric neutrino data. Atmospheric neutrinos cover a wide phase space in both energy and distance travelled, the parameters relevant for studying neutrino oscillations. We present here recent measurements of the three-flavor neutrino oscillation parameters using this atmospheric neutrino data, as well as new limits on mixing with a fourth sterile neutrino state.
\end{abstract}

\maketitle


\section{Introduction}

The \superk (\sk) detector has played an important role in the history of neutrino physics. The first definitive measurement of neutrino oscillations was made by \sk in 1998 in atmospheric neutrinos \cite{Fukuda:1998mi}. Since that time, the experiment has accumulated significantly more data, more than 11 live-years, and introduced a wide variety of neutrino event topologies into the oscillation analysis. In these proceedings we use these samples to address the significant issues in neutrino physics today: the detailed nature of oscillations among all three active neutrino flavors as well as the possibility of additional sterile neutrino states. 

The experiment also studies more distant neutrino sources. At lower energies, solar neutrinos are used to study oscillations driven by the smaller `solar' mass-splitting \cite{Abe:2010hy} and the searches for supernova neutrinos, both burst \cite{Ikeda:2007sa} and relic \cite{Bays:2011si}, are ongoing. At higher energies, we search for indirect evidence of dark matter through its annihilation into neutrinos \cite{Tanaka:2011uf}. We also search for evidence of nucleon decay \cite{Regis:2012sn,Nishino:2012ipa}. These proceedings will focus on oscillations in the atmospheric neutrino sample.

Atmospheric neutrinos are produced when cosmic rays collide with nuclei in the Earth's atmosphere, producing mesons which decay in flight producing muon and electron neutrinos. The cosmic rays arrive from all directions so the neutrinos produced are incident on our detector from all directions as well. We parameterize this direction as a `zenith angle,' \cz, defined as the cosine of the angle between the estimated neutrino direction and the vector pointing from the center of the detector to the center of the Earth. Downward-going neutrinos, which have a zenith angle near $1$, can travel as little as \val{10}{km} while upward-going neutrinos with \cz near $-1$ can travel as far as \val{13,000}{km}. Atmospheric neutrinos are also produced with a wide range of energies, ranging from less than a hundred MeV to more than a TeV.

\section{The \superk Detector}

\sk \cite{Fukuda:2002uc} is a large, underground, water-Cherenkov detector. It is arranged into two optically-separated regions: an inner-detector instrumented with 11,146 20-inch PMTs and an active-veto outer-detector instrumented with 1,885 8-inch PMTs, both filled with ultra-pure water. A fiducial volume is defined \val{200}{cm} inwards from the walls of the inner detector with a mass of \val{22.5}{kton}. \sk has had four run-periods. The SK-I, SK-III, and SK-IV had full 40\% photo-coverage while SK-II had only 19\% coverage due to an accident in 2001. The most recent period, SK-IV began with the installation of new front-end electronics. allowing trigger-less data acquisition into a buffer on which advanced software-based triggers are applied.

\begin{figure}
  \raisebox{-0.5\height}{\includegraphics[width=.25\textwidth]{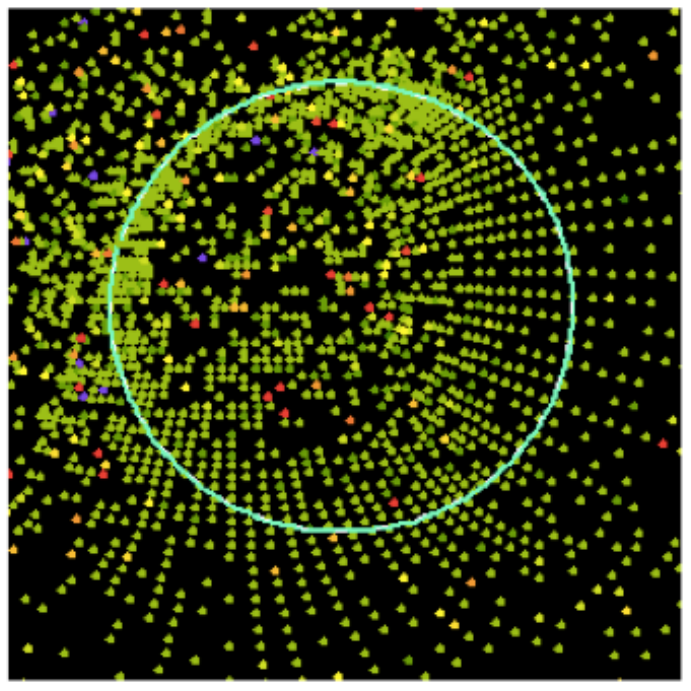}}
  \raisebox{-0.5\height}{\includegraphics[width=.45\textwidth]{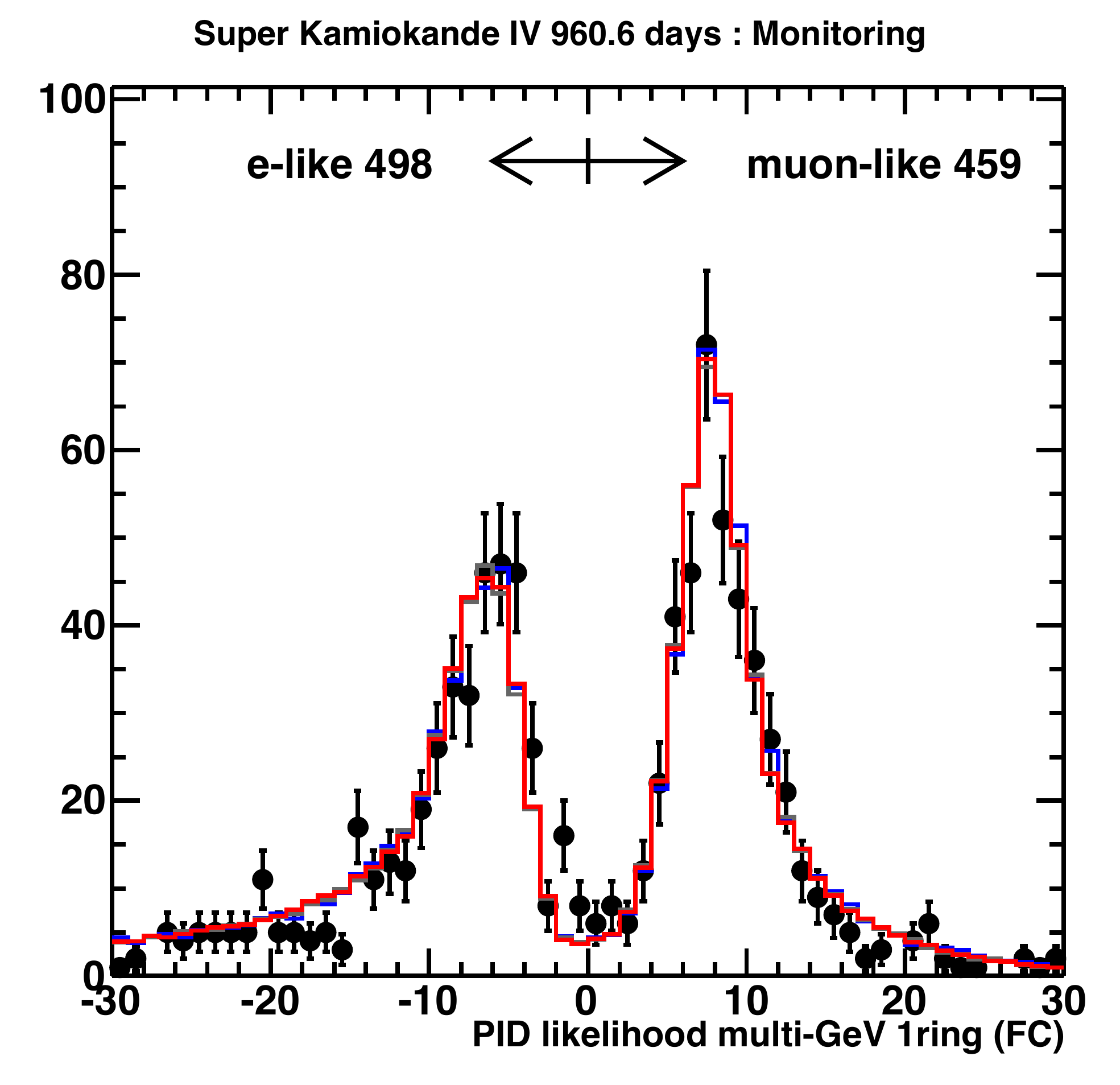}}
  \raisebox{-0.5\height}{\includegraphics[width=.25\textwidth]{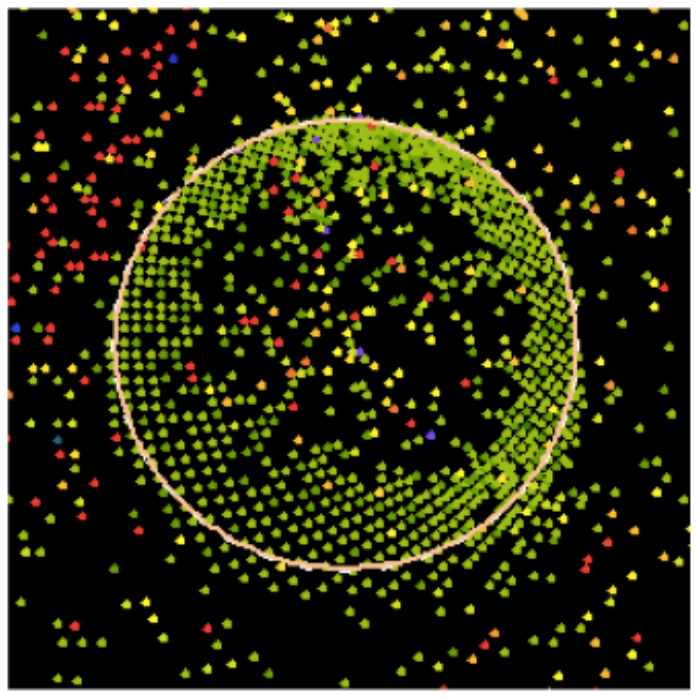}}
  \caption{The distribution of the likelihood parameter used to separate electron-like and muon-like rings with an example electron-like ring to the left and an example muon-like ring to the right.}
  \label{fig:pid}
\end{figure}

Neutrinos, of course, cannot be observed directly since they interact only via the weak nuclear force. Instead, we observe the charged particles produced when the neutrino interacts with a nucleus in the detector. The charged particles produced in these interactions are typically highly relativistic and will produce Cherenkov radiation when their kinetic energy is above a mass-dependent threshold energy (in water, \val{780}{keV} for electrons and \val{160}{MeV} for muons). Highly-relativistic particles will radiate Cherenkov photons in a cone (42\degrees in water for particles with velocity close to $c$) as long as the particle is above threshold, producing a ring of light projected onto the wall of the detector. The timing of the photon hits allows the vertex to be reconstructed, and direction of travel of the particle can be estimated from the vertex and the center of the ring. The more energetic the particle, the more light it will produce in the detector before falling below threshold. Particle-ID can also be determined based on the shape of the ring. Showering particles (electrons, photons) will produce many overlapping rings which appear as a single ring with a fuzzy edge. Non-showering particles (muons, pions, protons) travel in a consistent direction and produce a ring with a sharp outer boundary. A likelihood-based selection shown in \fref{pid} is used to separate these two ring types.

The neutrino oscillation probability depends on the initial neutrino flavor, the distance the neutrino travels, $L$, and the energy of that neutrino, $E$. The distance is estimated by extrapolating the direction of travel of the observed charged particles in the detector and extrapolating back to the atmosphere (above the detector or across the Earth). The energy is estimated by summing the energy seen in the detector and correcting for differences between showering and non-showering particles.

\begin{figure}
  \includegraphics[width=.45\textwidth]{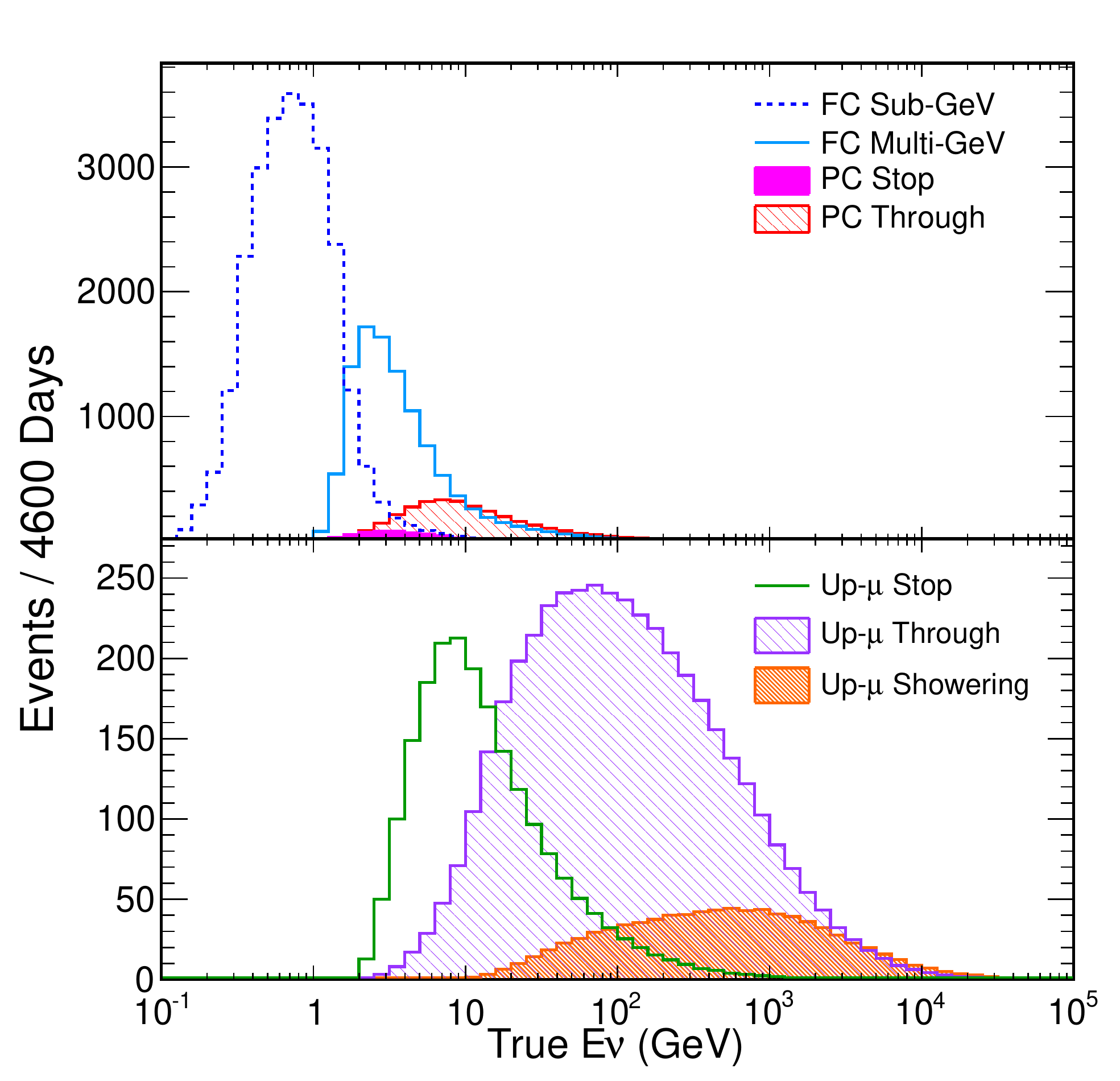}
  \caption{The true energy distribution from simulation of the fully-contained (sub-GeV and multi-GeV), partially-contained (stopping and through-going), and up-going muon samples (stopping, through-going non-showering, and through-going showering).}
  \label{fig:examples}
\end{figure}

There are three basic event topologies used in the atmospheric neutrino analysis which cover different neutrino energies (plotted in \fref{examples}). Fully-contained (FC) events have vertices inside the fiducial volume and stop before leaving the inner detector. They are the lowest-energy sample ranging from a few hundred MeV up to a few GeV. These events have the best energy resolution since all of the energy is contained and also have good PID information. Conversely, they also have the worst direction resolution since the outgoing lepton direction is less correlated with the incoming neutrino direction. In the oscillation analysis, these events are divided into 13 sub-samples. They are divided up by energy into sub-GeV and multi-GeV samples and binned either in lepton momentum (sub-GeV only) or both energy and \cz (sub- and multi-GeV). The samples are also divided up by number of decay electrons which can signify the presence of a charged pion produced below Cherenkov threshold and particle-id. Sub-GeV samples are split into $\mu$-like, $e$-like, and $\pi^0$-like (a neutral current-enhanced sample) while Multi-GeV samples are split into $\mu$-like, \nue-like and \nuebar-like. 

Partially contained (PC) events have vertices in the fiducial volume, but produce leptons that leave the inner detector. These events have long tracks and so are almost exclusively from \numu interactions and range in energy from a few GeV up to tens of GeV.  These events have better direction resolution than FC events due to their higher energy, but poorer energy resolution since the exiting muon carries some energy away out of the detector.  They are divided into stopping (which stop in the outer detector) and through-going which pass through the outer detector out into the rock. They are binned in both energy and \cz. 

Up-going muon (Up-$\mu$) events produce muons that start in the surrounding rock and then enter and pass through the outer detector into the inner detector. This sub-sample also starts at a few GeV but extends up to the highest energies in \sk. These events are only included if they are up-going where the bulk of the Earth has shielded the detector from the otherwise overwhelming cosmic-ray muon background. They are split into lower-energy stopping (stops in the inner detector) and higher-energy through-going (exits out the far side of the detector) sub-samples. The through-going events are further sub-divided into non-showering and showering. The critical energy at which energy loss by bremsstrahlung (the dominant process in showering) equals energy loss by ionization for muons is \val{900}{MeV} \cite{PDG} so evidence of showering provides an additional handle for estimating the energy of these muons which deposit a potentially large fraction of their energy in the rock around the detector. The showering up-$\mu$ sample is the highest-energy sample in \sk. The up-$\mu$ samples are binned only in \cz since the measured energy is only a rough lower bound on the initial neutrino energy.

\begin{figure}
  \includegraphics[width=.85\textwidth]{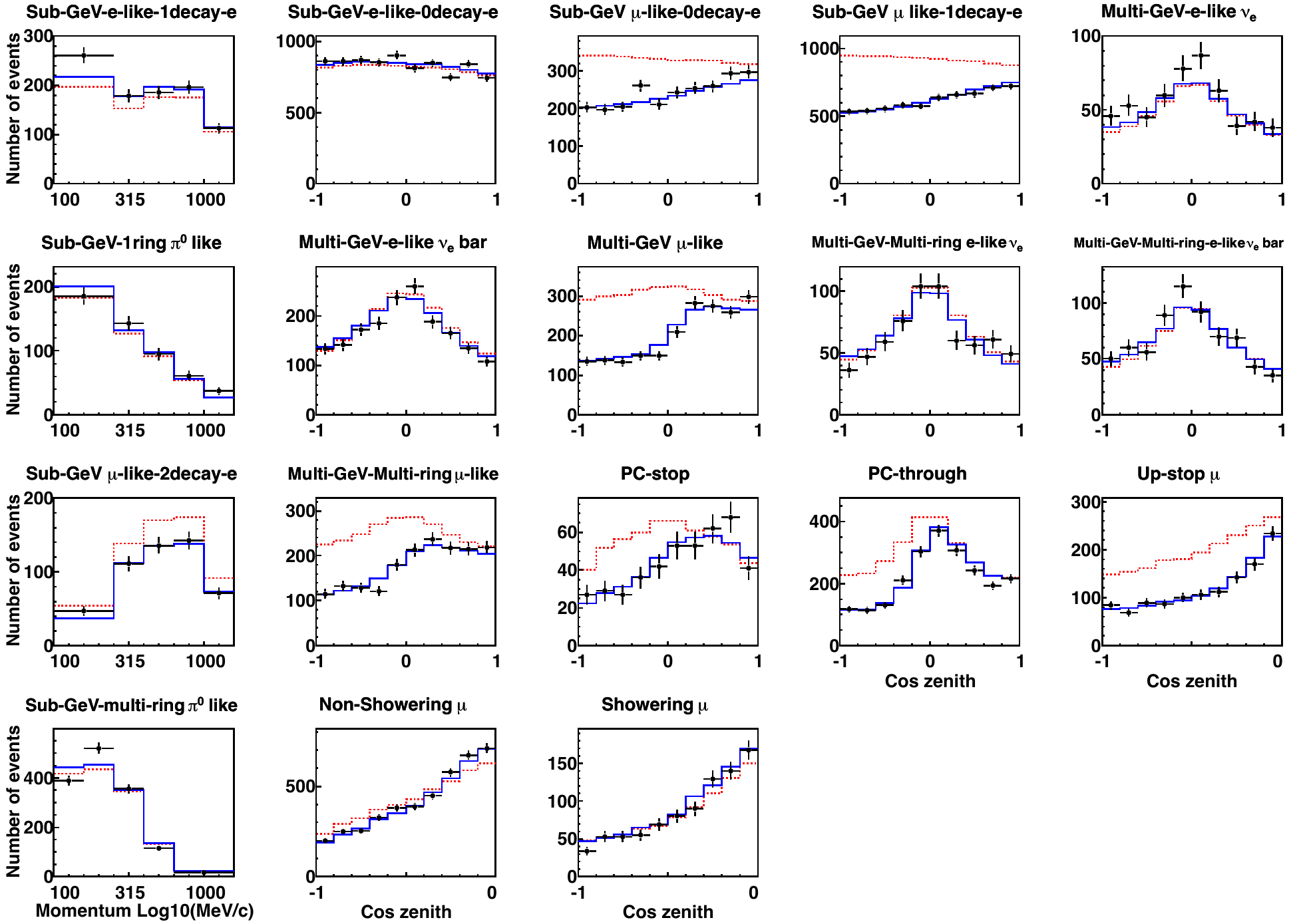}
  \caption{The samples data used in the oscillation analysis binned in zenith angle (or lepton momentum for sub-GeV samples without zenith binning in the left column). The black points show the SKI-IV data, the red dashed lines show the unoscillated MC and the blue line shows the best oscillation fit.}
  \label{fig:zenithdistros}
\end{figure}

These event samples, combined across all periods SKI-IV, are shown in \fref{zenithdistros}.

\section{Systematic Uncertainties}

The systematic uncertainties in the oscillation analyses fall into three broad categories: atmospheric neutrino flux, neutrino cross-sections, and detector effects. The flux uncertainties are calculated by the groups who calculate the flux prediction \cite{Honda:2011nf} and they are based on our knowledge of the cosmic ray flux to the Earth, our knowledge of the atmosphere, and our knowledge of the hadronic interactions that produce the mesons which decay into neutrinos. All three of these pieces are informed by external measurements, with the last being constrained by measurements in hadron production experiments (e.g. \cite{Catanesi:2008zzc}). In general, the largest uncertainties are in the absolute normalization of the flux. The ratios of various components to each other are known to much better precision -- for example electron/muon ratio, up-going/down-going ratio, etc.

The neutrino cross-section uncertainties are based on external cross-section measurements.  The detector uncertainties generally relate to energy scale and selection efficiency and vary between run periods (SKI-IV). These uncertainties are determined with laser and radioactive calibration sources and the naturally abundant cosmic-ray muon sample \cite{Ashie:2005ik,Wendell:2010md,Abe:2013gga}.

\section{Three-flavor Oscillation Analysis}

Oscillations of massive neutrinos has been well established in solar, reactor, and accelerator experiments \cite{Fukuda:1998mi,Fukuda:2001nj,Abe:2008aa}. These oscillations are typically parameterized with the PMNS mixing matrix which relates the neutrino mass and flavor eigenstates,
\begin{equation}
\left(\begin{array}{c}
\nue \\ 
\numu \\
\nutau
\end{array}\right)
= 
\left(\begin{array}{ccc}
 c_{12} & s_{12} & 0 \\
-s_{12} & c_{12} & 0 \\
      0 &      0 & 1
\end{array}\right)
\left(\begin{array}{ccc}
 c_{13} & 0 & s_{13}e^{-i\delta} \\
 0 & 1 & 0 \\
-s_{13}e^{i\delta} & 0 & c_{13} \\
\end{array}\right)
\left(\begin{array}{ccc}
1 &       0 &      0 \\
0 &  c_{23} & s_{23} \\
0 & -s_{23} & c_{23} 
\end{array}\right)
\left(\begin{array}{c}
\nu_1 \\ 
\nu_2 \\
\nu_3
\end{array}\right),
\end{equation}
where $s_{xy} = \sin\theta_{xy}$ and $c_{xy} = \cos\theta_{xy}$.  With the definitive measurement of a relatively large $\theta_{13}$ \cite{Abe:2011sj,An:2012eh}, we have entered a new era of neutrino oscillation measurements: we now know that all three neutrino flavors mix to a non-trivial degree. Given \sk's large range in $L$-over-$E$, oscillations among all three flavors can contribute. The major open questions now have shifted: is \sn{23} maximal? If not, is $\theta_{23} > 45\degrees$ or $< 45 \degrees$ (i.e. what `octant' is it in)? What is the sign of \dm{32}? 

\begin{figure}
  \includegraphics[width=.45\textwidth]{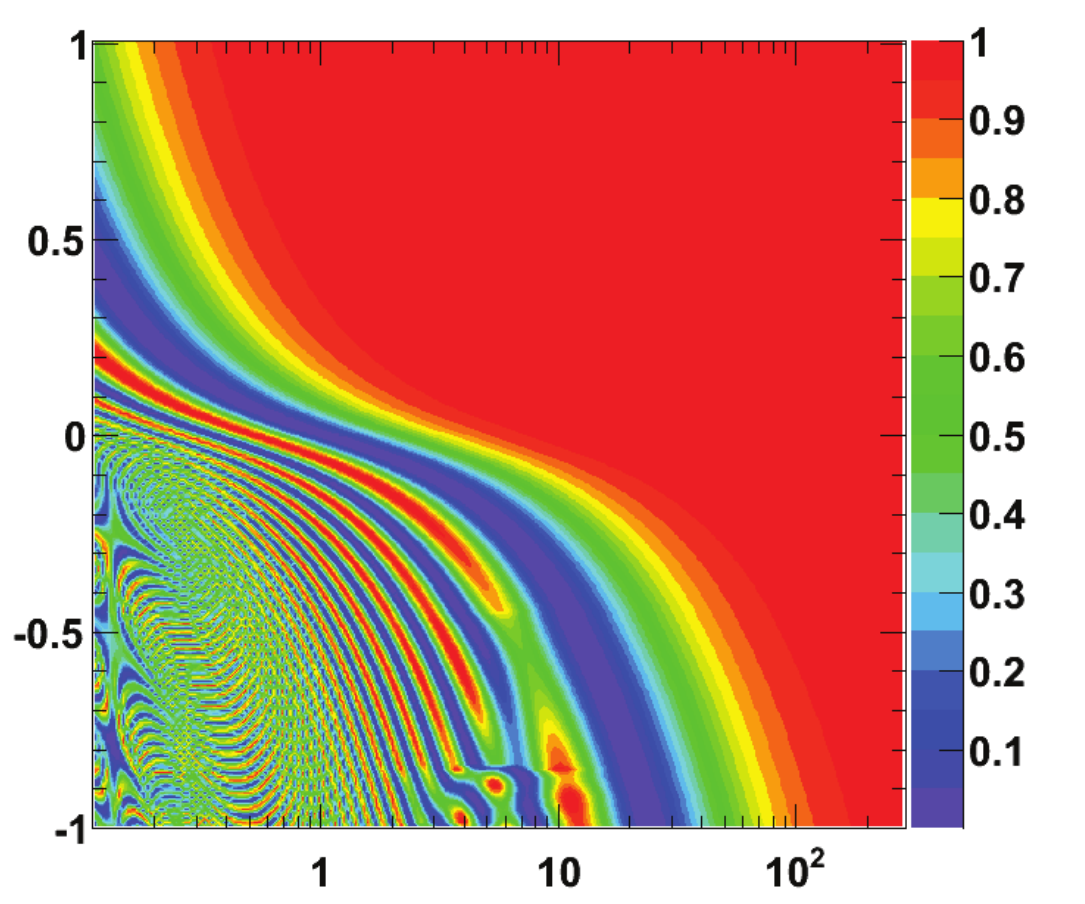}\includegraphics[width=.45\textwidth]{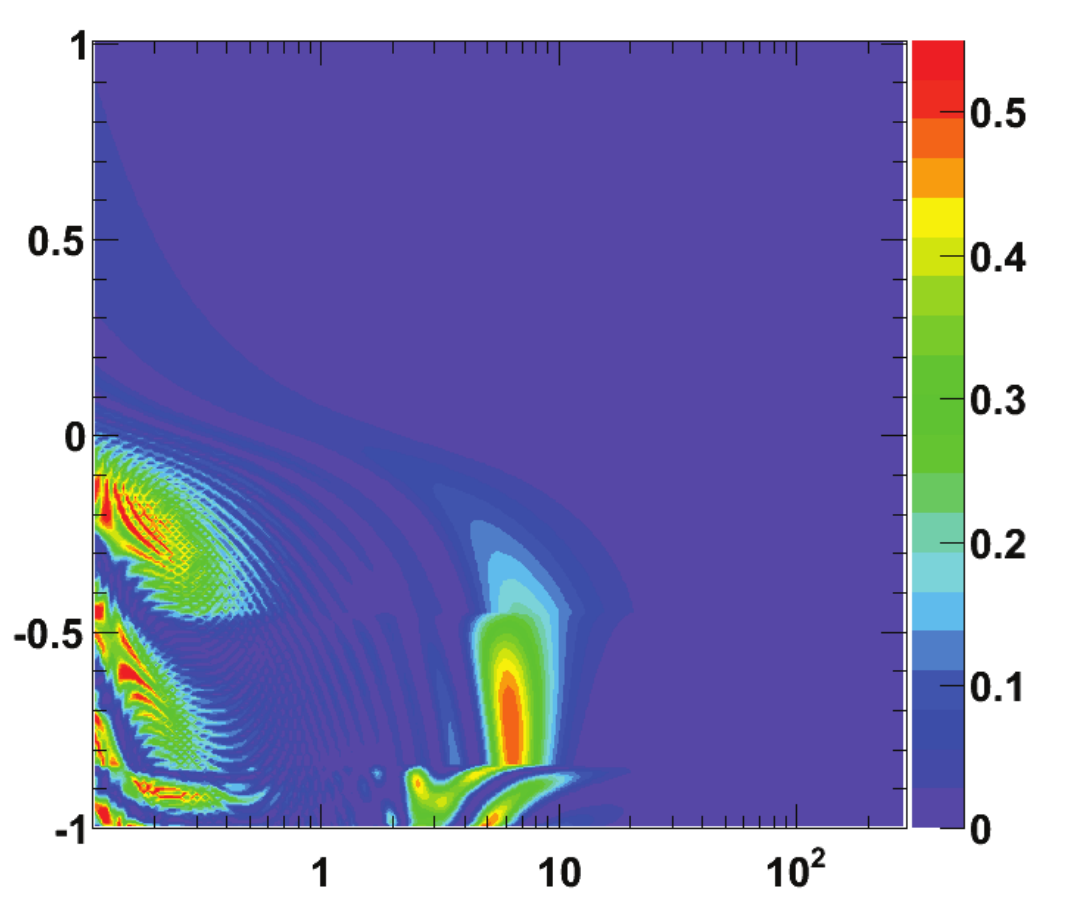}
  \caption{Oscillation probability of $\numu \to \numu$ on the left and $\numu \to \nue$ on the right.  The distortions around \val{10}{GeV} and \cz near -1 are due to matter effects.}
  \label{fig:oscillograms}
\end{figure}

In order to focus on these questions, the \sk zenith angle oscillation fit, which includes all oscillation parameters and matter effects following the prescription from \cite{Barger:1980tf} and described in detail in \cite{Wendell:2010md}, now includes a constraint on \snt{13} from \cite{PDG}. The oscillation probability in \sk is best shown in the two-dimensional plane of \cz vs. energy (\fref{oscillograms}). The distortions in the $\numu \to \numu$ oscillation bands and the enhancement of $\numu \to \nue$ around \val{10}{GeV} at $\cz < -0.5$ are driven by resonant matter effects as the neutrinos travel through the denser core of the Earth (the PREM model \cite{Dziewonski:1981xy} of the Earth density is used). In the normal hierarchy this resonant enhancement happens for neutrinos and in the inverted hierarchy it happens for antineutrinos, so the \nue and \nuebar enhanced samples described earlier enhance our sensitivity to the mass hierarchy. The $\numu \to \nue$ oscillation enhancements at low energy, driven by the smaller solar \dm{21}, are sensitive to \sn{23} and hence the $\theta_{23}$ octant (while $\numu \to \numu$ oscillations are primarily sensitive to \snt{23} and have little octant sensitivity). 

\begin{figure}
  \includegraphics[height=.25\textheight]{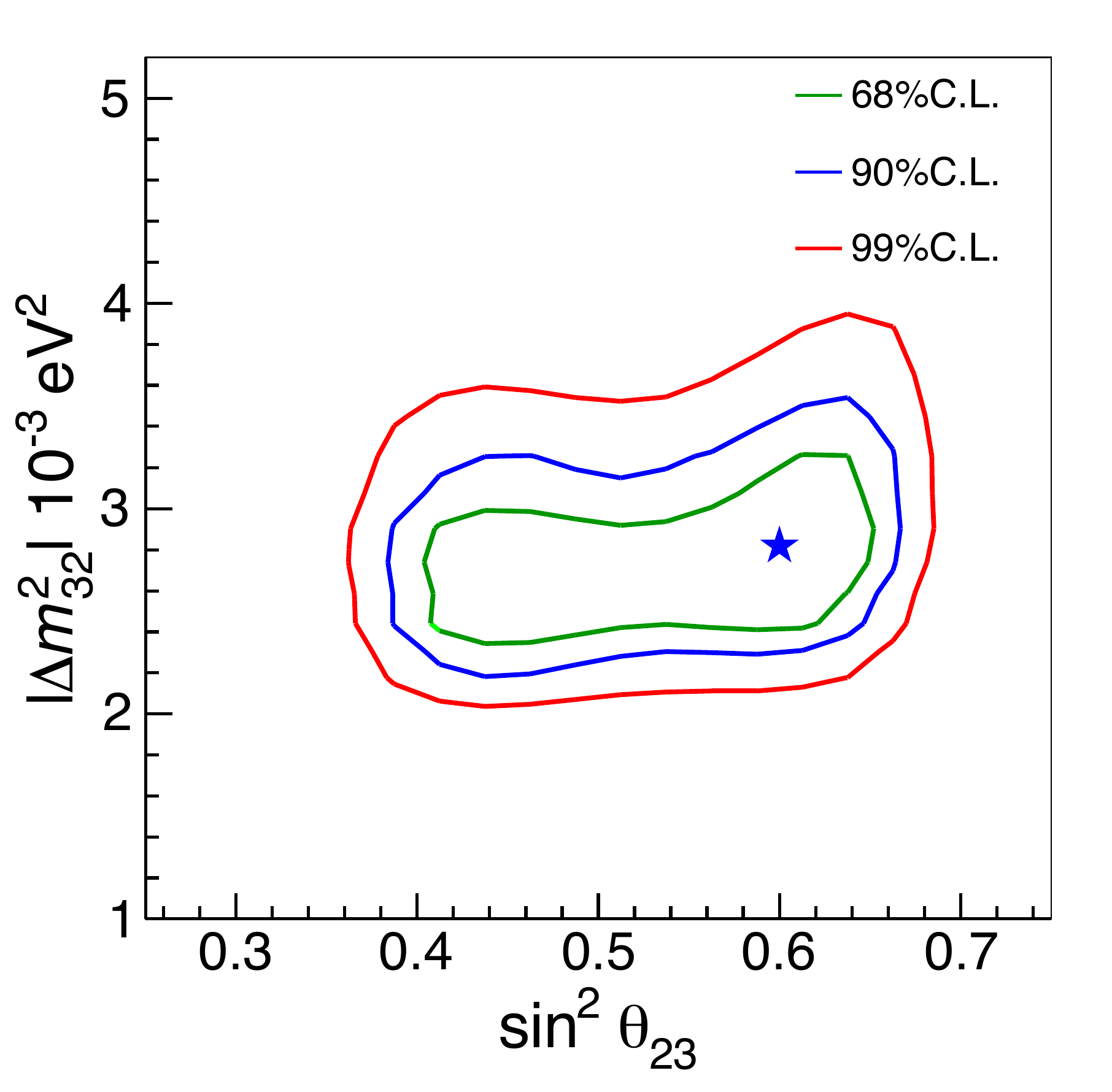}
  \includegraphics[height=.25\textheight]{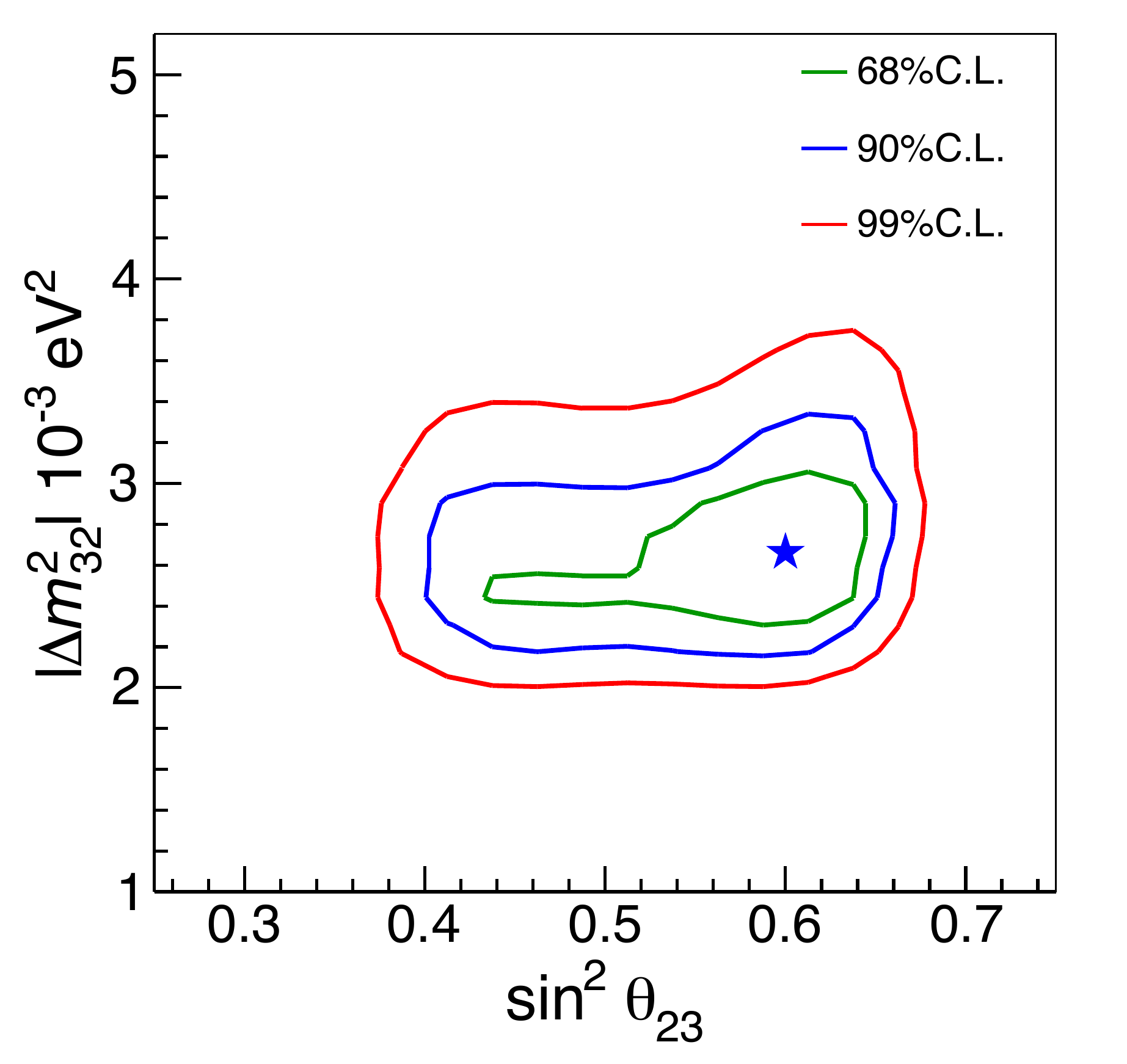}
  \includegraphics[height=.25\textheight]{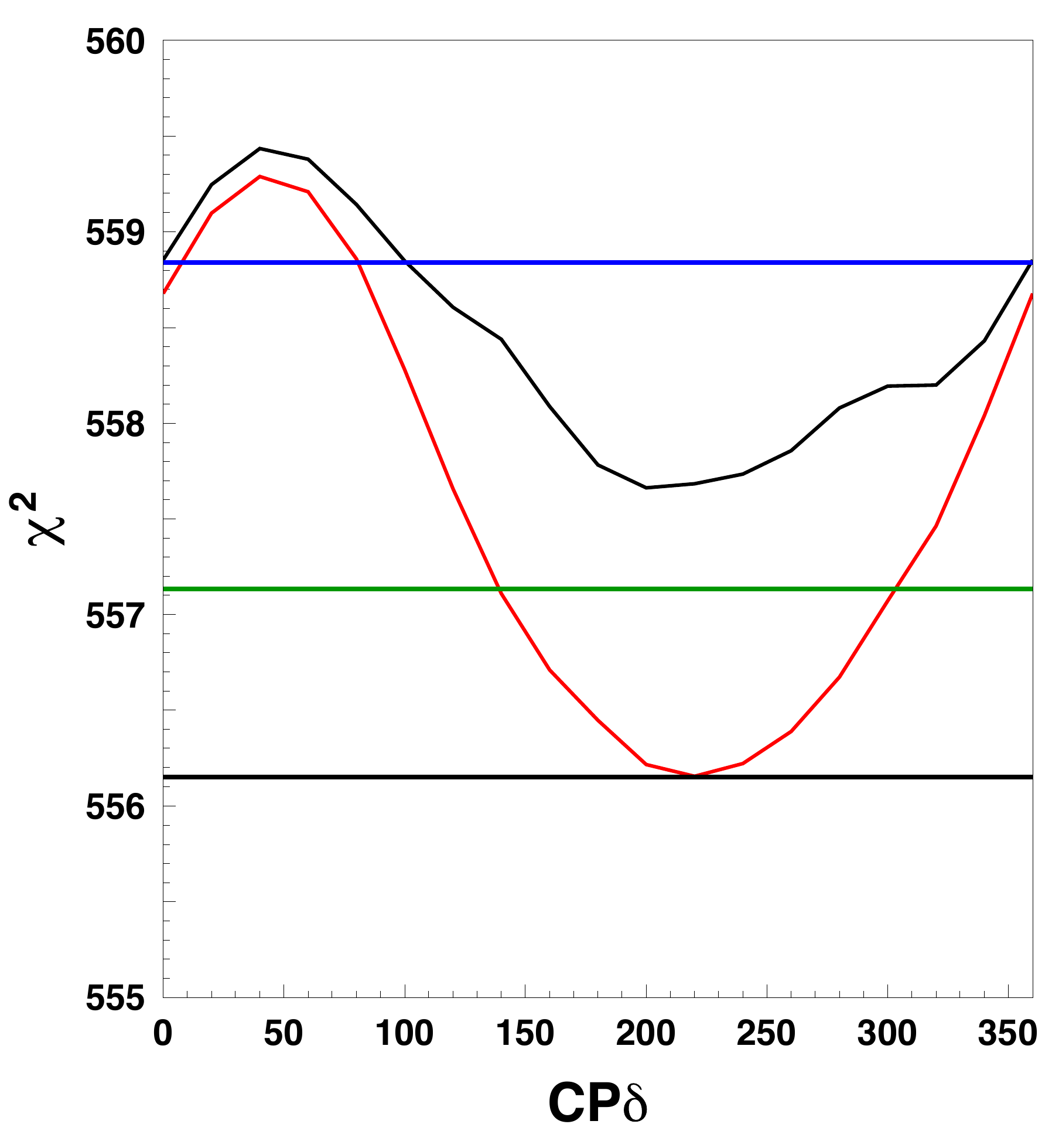}
  \caption{The left figure and center figure show the contours in \dm{32} vs. \sn{23} in the normal and inverted hierarchy assumptions, respectively. The figure on the right shows the $\chi^2$ vs. CP-$\delta$ for normal (black) and inverted (red). The green line represents $1\sigma$ and the blue line represents $2\sigma$ on this plot.}
  \label{fig:3nuresults}
\end{figure}

The results of the oscillation fit to SKI-IV data are shown in \fref{3nuresults} as contours in \dm{32} vs. \sn{32} separately for the normal ($\dm{32} > 0$) and inverted ($\dm{32} < 0$) mass hierarchy assumptions. Both fits weakly prefer the second octant, though more so with the inverted hierarchy assumption. Also shown is the $\chi^2$ distribution of both the normal and inverted hierarchy assumptions vs. the CP-violating $\delta$ parameter. The inverted hierarchy is preferred to the normal hierarchy by $1.2\sigma$ -- no definitive preference yet.  

\sk has also recently shown evidence of tau neutrino appearance consistent with three-flavor neutrino oscillations. An estimated $180.1 \pm 44.3\rm{(stat.)}^{+17.8}_{-15.2}\rm{(sys.)}$ tau neutrinos were observed in SKI-III, excluding the no-tau-appearance hypothesis by $3.8\sigma$ \cite{Abe:2012jj}.

\section{Sterile Neutrino Analysis}

A `sterile neutrino' is a hypothetical particle which does not interact via the weak nuclear force but does mix with the known neutrino states. Measurements of the width of the $Z^0$ boson at LEP have shown that exactly three light (less than half the $Z^0$-mass) neutrinos couple to the $Z^0$ and hence have weak interactions \cite{Decamp:1989tu}. Having three light neutrinos implies there can be only two independent mass-splittings. Two have been definitively observed in a range of experiments, the solar \dm{21} and the atmospheric \dm{32}. If a third independent mass-splitting is observed in neutrino oscillations, it necessarily implies a fourth neutrino state. Since it is already known that only three couple to the $Z^0$, this new fourth neutrino must be sterile.

When a fourth neutrino is added, the $3\times3$ mixing matrix must be expanded to a $4\times4$ mixing matrix adding 7 complex parameters. Unitarity reduces these 14 parameters to 5 independent ones: 3 new `angles' and 2 new phases. The angles can be parameterized multiple ways, but in these proceedings we have chosen to use the magnitudes of the matrix elements: \uesq, \umsq, and \utsq. This type of model, with a fourth sterile neutrino widely separated in mass from the three active neutrinos, is often called `3+1.' Additional neutrinos can be added to make `3+2' or `1+3+1' models, adding 8 additional free parameters to the oscillation model.

Hints of possible sterile neutrinos have appeared in several experiments \cite{Abazajian:2012ys}. First the LSND experiment and then the MiniBooNE experiment saw hints of $\numu \to \nue$ oscillations consistent with two-flavor oscillations with a $\dmnew \approx \val{1}{eV^2}$ \cite{Aguilar-Arevalo:2013pmq,Athanassopoulos:1997pv}. The effective two-flavor angle can be translated into the 3+1 parameterization above as $\snt{\mu e} = \uesq\umsq$.
Some hints of sterile oscillations have also been seen in lower-than-expected rates in short-baseline reactor experiments and in the radioactive source calibrations of gallium solar neutrino experiments \cite{Kopp:2013vaa}. If these low rates are interpreted as two-flavor oscillations, they are consistent with $\dmnew > ~\val{1}{eV^2}$ and with $\uesq \approx 0.02$ \cite{Kopp:2013vaa}. The two-flavor angle in this case is, $\snt{ee} = \uesq \left(1 - \uesq \right)$.

\sk's sensitivity to sterile neutrinos derives primarily from its observation of muon neutrinos across a wide range of $L$-over-$E$. Muon neutrino disappearance measurements at both short and long baselines are sensitive to oscillations via \umsq, driven by \dmnew. At short baseline this can approximate two-neutrino mixing (the formula is analogous to \snt{ee}), but at long baselines like in \sk, the signature of non-zero \umsq is fast oscillations away from normal three-flavor oscillations. Long-baseline measurements like \sk are also sensitive to sterile oscillations driven by \dm{32}, but $\numu \to \nu_{s}$ instead of $\numu \to \nutau$. These oscillations require non-zero \umsq and \utsq and also introduce a new kind of matter effect. Where the more well known charged current (CC) matter effects arise from the potential difference between \nue's with CC and neutral current (NC) interactions and \numu and \nutau with only NC interactions, these new NC matter effects arise from the potential difference between \nue, \numu, and \nutau which have NC interactions and $\nu_s$ which has no weak interactions. These matter effects tend to shift around the effective \dm{32} in $\numu$ disappearance and distort the shape of the oscillations at the longest lengths through the Earth.

Unfortunately, it is too computationally difficult to use the complete, fully generic 3+1 sterile model in the oscillation fit. Some simplifying assumptions are required (based on the model used in \cite{Maltoni:2007zf}), the first being that the model is only 3+1. One of the significant advantages of the fit to atmospheric neutrinos is that 3+1 and 3+N models look the same to first order, so a simpler fit can constrain more complex models. We also assume there are no sterile-\nue oscillations ($\uesq \to 0$), which is reasonable since the constraints from the \nue disappearance analyses limit the possible size of such an effect to the point where it becomes impossible to observe in \sk \cite{Maltoni:2007zf}. Oscillations are assumed to be fast, a good assumption for all \dmnew values in low-energy samples, but one that starts to break down below \val{0.8}{eV^2} in the highest-energy samples. 

The final difficulty in the model is that CC and NC matter effects cannot be efficiently calculated simultaneously because there is no analytic diagonalization of the Hamiltonian with more than one non-zero element in the matter potential. So, two different fits are performed. One fit is performed for only the parameter \umsq with just the CC matter effects -- \umsq is over-constrained if the standard matter effects are not included. A second fit is performed for \umsq and \utsq which includes only the NC matter effects. Since \utsq sets the size of these matter effects, their observation or non-observation provides a strong constraint on \utsq. Effectively, the CC matter effects fit sets the most accurate limit on \umsq and the NC matter effects fit sets the limit on \utsq (even though \umsq is included in both fits). Note that in the latter fit there is a unitarity bound keeping $\umsq+\utsq \leq 1$

Neither fit finds any evidence of sterile neutrino oscillations. \umsq is constrained to $< 0.034$ and \utsq to $< 0.28$ at the 99\% confidence level.  These new \sk limits can be compared against other \numu disappearance measurements in \fref{sterileresults}. Our results provide new constraints away from the peak sensitivity region of the short-baseline measurements of \umsq as well as the strongest constraint on \utsq mixing, meaning that we strongly favor $\numu \to \nutau$ over $\numu \to \nu_s$ oscillations.

\begin{figure}
  \includegraphics[height=.25\textheight]{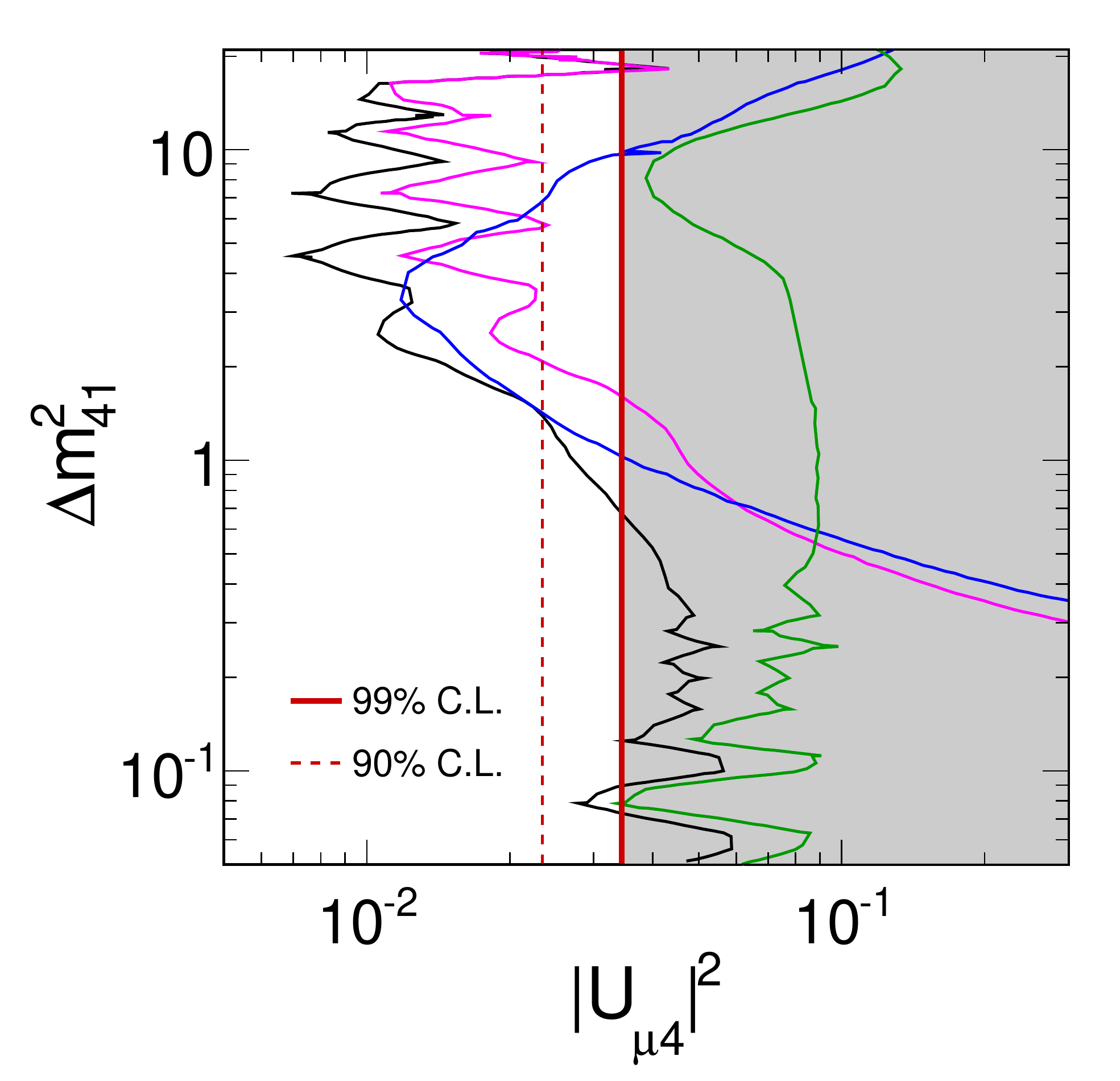}
  \includegraphics[height=.25\textheight]{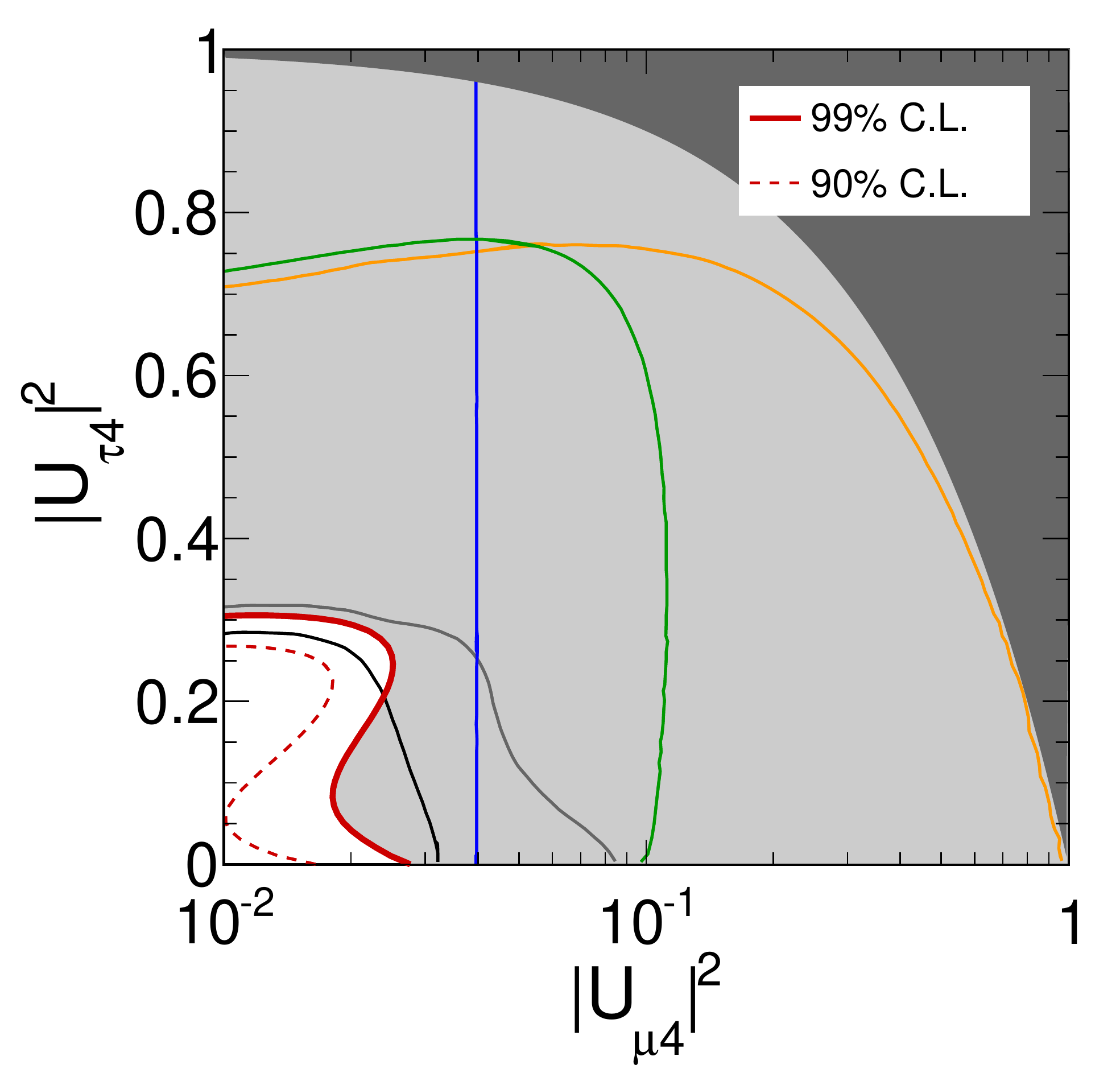}
  \caption{99\% limits (90\% limit dashed) from \sk in red compared against limits from various other experiments, taken from \cite{Kopp:2013vaa}. MINOS is in green, CDHS is in blue, MiniBooNE is in magenta, solar experiments in yellow, and a global fit in black. The light gray region is excluded by the \sk measurement and the dark grey region on the right is excluded by unitarity.}
  \label{fig:sterileresults}
\end{figure}

\section{Conclusion}

\sk has accumulated more than 11 live-years of atmospheric neutrino data. With this data we have made precise measurements of \dm{32} and \sn{23} in a three-flavor oscillation framework. The data favor the inverted mass-hierarchy by a little more than 1 unit of $\Delta \chi^2$, not enough to be significant. We have also seen no evidence of sterile neutrino mixing in atmospheric neutrinos, placing new limits on sterile mixing parameters that are independent of the size of \dmnew and the number of additional sterile states beyond one.


\begin{theacknowledgments}
We gratefully acknowledge the cooperation of the Kamioka Mining and Smelting Company. \sk has been built and operated from funds provided by the Japanese Ministry of Education, Culture, Sports, Science and Technology, the U.S. Department of Energy, and the U.S. National Science Foundation. This work was partially supported by the  Research Foundation of Korea (BK21 and KNRC), the Korean Ministry of Science and Technology, the National Science Foundation of China, and the Spanish Ministry of Science and Innovation (Grants FPA2009-13697-C04-02 and Consolider-Ingenio-2010/CPAN).
\end{theacknowledgments}

\bibliographystyle{aipproc}   

\bibliography{ppc}

\end{document}